\documentclass[aps,reprint,twocolumn,groupedaddress,showpacs]{revtex4}
\usepackage{graphicx}
\usepackage{color}
\usepackage{amsmath}
\usepackage{txfonts}
\usepackage{bm}

\usepackage{ulem}

\newcommand{\red}{\textcolor{red}}

\begin{document}

\title{
Energetics of synchronization in coupled oscillators rotating on circular trajectories
}

\author{Yuki Izumida$^1$}
\thanks{Present address: Department of Complex Systems Science, Graduate School of Information Science, Nagoya University, Nagoya 464-8601, Japan}
\email{izumida@is.nagoya-u.ac.jp}
\author{Hiroshi Kori$^1$}
\author{Udo Seifert$^2$}
\affiliation{$^1$Department of Information Sciences, 
Ochanomizu University, Tokyo 112-8610, Japan\\
$^2$II. Institut f\"ur Theoretische Physik, Universit\"at Stuttgart, 70550 Stuttgart, Germany}


\begin{abstract}
We derive a concise and general expression of the energy dissipation rate for coupled oscillators rotating on circular trajectories by 
unifying the nonequilibrium aspects with the nonlinear dynamics via stochastic thermodynamics.
In the framework of phase oscillator models, it is known that the even and odd parts of the coupling function express the effect on collective and relative dynamics, respectively.
We reveal that the odd part always decreases the dissipation upon synchronization, while the even part yields a characteristic square-root change of the dissipation near the bifurcation point whose sign depends on the specific system parameters.
We apply our theory to hydrodynamically coupled Stokes spheres rotating on circular trajectories that can be 
interpreted as a simple model of synchronization
of coupled oscillators in a biophysical system. We show that the coupled Stokes spheres gain the ability to do more work on the surrounding fluid 
as the degree of phase synchronization increases.
\end{abstract}

\maketitle

\section{Introduction}
Coupled oscillators and their
synchronization phenomena are ubiquitously found in a variety of
scientific and engineering fields~\cite{W,K,PRK}. They are typical
examples of nonequilibrium dissipative systems that are maintained by a
balance of energy injection and dissipation. The relationship between
synchronization and energy dissipation has been attracting much
interest, e.g., in the context of low Reynolds-number
hydrodynamics~\cite{RS,VJ,KN,EL,OV,JG} since Taylor's classical work on
hydrodynamic synchronization of active objects 
with periodic motions~\cite{T}. 
Recent extensive theoretical and experimental studies
on beating eukaryotic flagella and cilia have elucidated the underlying
physical mechanism of hydrodynamic synchronization based on a simplified phase-description without losing its
essence~\cite{VJ,RL,UG,UG2011prl,UG2012EPJE,GPT,GPT2,MKRJF,BWPG,KDBBPSHC,UG2011SM}.
In this phase-description, they are simply modeled as coupled oscillators whose periodic motions are described by phase
equations. These tiny oscillators are motive-powered by a collection of
molecular machines that convert chemical energy into mechanical work
in a noisy thermal environment~\cite{PKT}. The hydrodynamic flow generated by such beating flagella and cilia plays a vital
and versatile role in living organisms, utilized, e.g., in the motility of
sperm and material transport by metachronal waves~\cite{LP,PKT,UG2011SM}. One important aspect is to understand how
synchronization and desynchronization between the oscillators that operate in a noisy environment affect energy dissipation. 
To develop this energetics of synchronization, we need to unify energetic concepts usually treated
in thermodynamics with the theory of coupled oscillators usually treated
in nonlinear dynamics.  Such a unification from the
stochastic thermodynamics point of view~\cite{ks,us} has been developed, in the
analysis of collective dynamics based on a nonequilibrium
equality~\cite{SS} and in the optimization of the energy-conversion efficiency in all-to-all coupled
many-oscillators systems~\cite{AI}.

In the present paper, we study the relationship between synchronization and energy dissipation rate
for the simplest and prototypical case of coupled oscillators
rotating on circular trajectories described by phase equations via stochastic thermodynamics.
While the difficulty of estimating the energy dissipation rate
comes from the fact that these oscillators cannot be treated independently,
we nevertheless can derive a concise and
general expression of the energy dissipation rate for them, which can be applied to any type of weak coupling.
Our expression elucidates the relationship between synchronization and energy dissipation rate, 
where the decomposition of the coupling function into the even and odd parts, 
which express the effect on collective and relative dynamics, respectively, has a key role. From this decomposition, we reveal that the contribution from the odd part always decreases the dissipation upon  frequency synchronization, while the contribution from the even part yields a characteristic square-root change of the dissipation near the bifurcation point whose sign depends on the specific system parameters.
We demonstrate our theory by applying it to a model of two rotating Stokes
spheres on circular trajectories synchronized through hydrodynamic coupling~\cite{UG2011prl,UG2012EPJE,KDBBPSHC}.

The organization of the rest of the present paper is as follows. In Sec.~\ref{Model}, we introduce our model of coupled oscillators on circular trajectories 
described by phase equations.
In Sec.~\ref{Energy dissipation rate and synchronization}, we derive the expression of the energy dissipation rate for our model as the main result. In Sec.~\ref{Example}, we apply our expression to the example of hydrodynamic synchronization 
of coupled Stokes spheres. We summarize our paper in Sec.~\ref{Summary and discussion}.

\section{Model}\label{Model}
\begin{figure}
\includegraphics[scale=0.45]{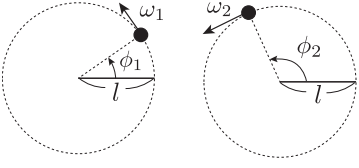} \caption{Schematic illustration of the model.}\label{machine}
\end{figure}
We consider two oscillators immersed in a viscous fluid as a
thermal environment, where the position of each oscillator is
constrained on a circle with radius $l$ on the same plane
(Fig.~\ref{machine}). 
Both are coupled, e.g.,
hydrodynamically~\cite{UG2011prl,UG2012EPJE,KDBBPSHC} or magnetically~\cite{MLBBS,LMBBS}.
We assume that their dynamical behavior can be described by the following phase equations
\begin{eqnarray}
&&\dot{\phi}_1=\omega_1+K\Gamma(\phi_1-\phi_2)+\zeta_1,\label{eq.phase_eq_1}\\
&&\dot{\phi}_2=\omega_2+K\Gamma(\phi_2-\phi_1)+\zeta_2.\label{eq.phase_eq_2}
\end{eqnarray}
Here, $\phi_i$ ($i=1,2$) is the phase of the $i$-th oscillator that
increases counterclockwise, and $\dot{\phi}_i \equiv \frac{d\phi_i}{dt}$. Because $l\phi_i$ denotes the arc length of
the circle measured from an origin to the oscillator, the kinematic velocity of the oscillator in our model is always proportional to the phase velocity as $l\dot{\phi}_i$. 
$\omega_i$ is the natural frequency of the $i$-th oscillator, which may be interpreted as resulting from a
driving force. $\Gamma$ is a $2\pi$-periodic coupling function between the oscillators
and $K >0$ is a coupling strength, respectively.  $\zeta_i$ is Gaussian white noise
whose correlation function obeys $\left<\zeta_i(t)\right>=0$ and 
$\left<\zeta_i(t)\zeta_j(t')\right>=2{\tilde{D}} \delta_{ij}\delta(t-t')$,
where $\tilde D\equiv Dl^{-2}$ is the normalized diffusion coefficient $D\equiv \mu \epsilon$
with $\mu$ and $\epsilon\equiv k_{\rm B}T$ being the constant mobility and the
noise intensity of the thermal environment where $k_{\rm B}$ and $T$
denote the Boltzmann constant and the temperature, respectively.
Hereafter $\left<\cdot \right>$ denotes a noise average. 
In general, phase equations may be more complicated, where the natural frequency is phase-dependent and the coupling function is no longer the function of the phase difference
(see Eqs.~(\ref{eq.oseen_1}) and (\ref{eq.oseen_2}) below as an example).
However, applying standard techniques in nonlinear dynamics such as cycle averaging under a
suitable variable transformation, we can reduce such phase
equations into the same form as Eqs.~(\ref{eq.phase_eq_1}) and
(\ref{eq.phase_eq_2}) to a good approximation
as long as the coupling strength is sufficiently
weak~\cite{K}, allowing us to discuss the general aspects of the energetics
of synchronization in coupled oscillators on circular trajectories.

By taking the average or the difference between Eqs.~(\ref{eq.phase_eq_1}) and (\ref{eq.phase_eq_2}), we obtain
\begin{eqnarray}
&&\frac{\dot{\phi}_1+\dot{\phi}_2}{2}=\bar{\omega}+K\Gamma_{\rm e}(\phi_1-\phi_2)+\frac{\zeta_1+\zeta_2}{2},\label{eq.collective}\\
&&\dot{\phi}_1-\dot{\phi}_2=\Delta \omega+2K\Gamma_{\rm o}(\phi_1-\phi_2)+\zeta_1-\zeta_2,\label{eq.relative}
\end{eqnarray}
respectively, where $\bar{\omega}\equiv \frac{\omega_1+\omega_2}{2}$ and $\Delta \omega \equiv \omega_1-\omega_2$.
Here, $\Gamma_{\rm e}$ and $\Gamma_{\rm o}$ are the even and odd parts of the coupling function defined as
\begin{eqnarray}
\Gamma_{\rm e}(\phi_1-\phi_2)\equiv \frac{\Gamma(\phi_1-\phi_2)+\Gamma(-(\phi_1-\phi_2))}{2},\label{eq.even}\\
\Gamma_{\rm o}(\phi_1-\phi_2)\equiv \frac{\Gamma(\phi_1-\phi_2)-\Gamma(-(\phi_1-\phi_2))}{2}.\label{eq.odd}
\end{eqnarray}
From Eqs.~(\ref{eq.collective}) and (\ref{eq.relative}), the even and odd parts of the coupling function express the effect on collective and relative dynamics, respectively.
Because $\Gamma_{\rm o}$ has a potential function
\begin{eqnarray}
U(\phi_1-\phi_2)\equiv -\int^{\phi_1-\phi_2} \Gamma_{\rm o}(\theta')d\theta',
\end{eqnarray}
Eqs.~(\ref{eq.phase_eq_1}) and
(\ref{eq.phase_eq_2}) can be rewritten as
\begin{eqnarray}
\dot{\phi_i}=\omega_i+K\Gamma_{\rm e}(\phi_1-\phi_2)-K\frac{\partial U}{\partial \phi_i}+\zeta_i\equiv \mu {\mathcal F}_i+\zeta_i.\label{eq.phase_eq}
\end{eqnarray}

In the absence of noise ($\epsilon=0$), 
the condition for frequency synchronization 
\begin{eqnarray}
\dot{\phi}_1-\dot{\phi}_2=\Delta \omega+2K \Gamma_{\rm o}(\phi_1-\phi_2)=0\label{eq.def_freq_synchro}
\end{eqnarray}
is equivalent to the existence of a phase-locked solution 
$\phi_i=\Omega t+\phi_i^0$ for Eq.~(\ref{eq.phase_eq}) where $\Omega$ and $\phi_i^0$ are constants denoting the shared frequency and the phase offset, respectively.
This condition for frequency synchronization is met if $K$ and $\Delta \omega$ satisfy
\begin{eqnarray}
-2K\Gamma_{\rm o, max}\le \Delta \omega \le -2K\Gamma_{\rm o, min}, \label{eq.bifurcation_range}
\end{eqnarray}
where 
$\Gamma_{\rm o,min}$ and $\Gamma_{\rm o, max}$ denote the minimum and the maximum values of $\Gamma_{\rm o}$, respectively~\cite{note1}.  
At the equalities of Eq.~(\ref{eq.bifurcation_range}), 
the phase-locked solution vanishes via
a saddle-node bifurcation, and phase slips
periodically occur in parameter ranges that do not
satisfy Eq.~(\ref{eq.bifurcation_range}) leading to desynchronization.

In the presence of noise ($\epsilon \ne 0$), the frequency
synchronization no longer exists in a strict sense.
However, for sufficiently weak noise, when Eq.~(\ref{eq.bifurcation_range}) is satisfied,
Eq.~(\ref{eq.def_freq_synchro}) approximately holds, 
so that the concept of synchronization is still meaningful. 
Under this assumption, we can expect that the system satisfying Eq.~(\ref{eq.bifurcation_range}) stays in the vicinity of one of the stable phase-locked solutions.

The Fokker-Planck equation corresponding to Eq.~(\ref{eq.phase_eq}) is given by
\begin{eqnarray}
\frac{\partial p(\phi_1, \phi_2, t)}{\partial t}=-\sum_{i=1}^2 \frac{\partial {\mathcal J}_i(\phi_1, \phi_2, t)}{\partial \phi_i},\label{fp.eq}
\end{eqnarray}
where we denote by $p(\phi_1,\phi_2,t)$ the probability distribution of the phases of the oscillators 
and by ${\mathcal J}_i(\phi_1, \phi_2, t)$ the probability current defined as
\begin{eqnarray}
{\mathcal J}_i(\phi_1, \phi_2, t)\equiv \mu {\mathcal F}_i p(\phi_1, \phi_2, t)-\tilde{D} \frac{\partial p(\phi_1, \phi_2, t)}{\partial \phi_i}.\label{eq.prob_current}
\end{eqnarray}
The stationary solution $p^{\rm ss}(\phi_1,\phi_2)$ satisfies $\frac{\partial p(\phi_1,\phi_2,t)}{\partial t}=0$.  
We can then define the mean frequency $\Omega_i$ as 
\begin{eqnarray}
\Omega_i\equiv \left<\dot{\phi}_i\right>=\int_0^{2\pi}d\phi_1 \int_{0}^{2\pi}d\phi_2 {\mathcal J}_i^{\rm ss}(\phi_1, \phi_2), \label{eq.def_Omega}
\end{eqnarray}
by using the stationary probability current ${\mathcal J}_i^{\rm ss}(\phi_1, \phi_2)$~\cite{us}.
A formal expression of $\Omega_i$ can be obtained as follows.
The probability distribution $f(\theta,t)$ of the phase difference $\theta \equiv \phi_1-\phi_2$
is governed by the Fokker-Planck equation
\begin{eqnarray}
&&\frac{\partial f(\theta, t)}{\partial t}=-\frac{\partial {\mathcal J}(\theta,t)}{\partial \theta},\label{eq.theta}\\
&&{\mathcal J}(\theta,t)\equiv \left(\Delta \omega+2K\Gamma_{\rm o}(\theta) \right)f(\theta, t)-2\tilde{D} \frac{\partial f(\theta, t)}{\partial \theta},\label{eq.j_f}
\end{eqnarray}
where 
the periodic boundary condition $f(\theta+2\pi,t)=f(\theta,t)$ is imposed. 
By putting $\frac{\partial f(\theta,t)}{\partial t}=0$ in Eq.~(\ref{eq.theta}), 
we obtain the following expression for the stationary distribution $f^{\rm ss}(\theta)$~\cite{S1,S2,PRK,R}:
\begin{eqnarray}
f^{\rm ss}(\theta)=\mathcal{N} \exp \left(-\frac{U_{\rm eff}(\theta)}{2\tilde{D}}\right)\int_{\theta}^{\theta+2\pi}dy \exp \left(\frac{U_{\rm eff}(y)}{2\tilde{D}}\right),\label{eq.distri_f}
\end{eqnarray}
where $U_{\rm eff}(\theta)\equiv 2KU(\theta)-\Delta \omega\theta$,
and
\begin{eqnarray}
\mathcal{N}\equiv \left[\int_{0}^{2\pi}d\theta \int_{\theta}^{\theta+2\pi}dy \ \exp \left(\frac{U_{\rm eff}(y)-U_{\rm eff}(\theta)}{2\tilde{D}}\right) \right]^{-1}.\label{eq.normalize}
\end{eqnarray}
By using Eqs.~(\ref{eq.collective}), (\ref{eq.relative}), (\ref{eq.def_Omega}), and (\ref{eq.distri_f}), we easily obtain the averaged mean-frequency $\bar{\Omega}$ and the averaged frequency-difference $\Delta {\Omega}$:
\begin{eqnarray}
\bar{\Omega}&&\equiv  \frac{\Omega_1+\Omega_2}{2}=\bar{\omega}+K\left<\Gamma_{\rm e}(\theta)\right>,\label{eq.omega_center}\\
\Delta \Omega &&\equiv \Omega_1-\Omega_2=\Delta \omega+2K\left<\Gamma_{\rm o}(\theta)\right>\nonumber\\
&&=4\pi \tilde{D} \mathcal{N} \left[1-\exp \left(-\frac{\Delta \omega \pi}{\tilde{D}}\right)\right]\label{eq.omega_relative},
\end{eqnarray}
where $\left<\Gamma_{\rm e, o}(\theta)\right>\equiv \int_0^{2\pi}\Gamma_{\rm
e, o}(\theta)f^{\rm ss}(\theta)d\theta$. Using Eqs.~(\ref{eq.omega_center})
and (\ref{eq.omega_relative}), we can also obtain an explicit expression
for each mean frequency $\Omega_i$. 
In the limit of $K \to 0$, $\mathcal{N}\to \left[-\frac{4\pi \tilde{D}}{\Delta
\omega}\left(\exp \left(-\frac{\Delta \omega
\pi}{\tilde{D}}\right)-1\right)\right]^{-1}$ and $\Omega_i\to \omega_i$, implying desynchronization, 
while for sufficiently large $K$ that satisfies Eq.~(\ref{eq.bifurcation_range}),
$\mathcal{N}\to 0$ and $\Omega_i \to \bar{\omega}+K\left<\Gamma_{\rm
e}(\theta)\right>$, implying synchronization.

\section{Energy dissipation rate and synchronization}\label{Energy dissipation rate and synchronization}
\subsection{Expression of energy dissipation rate under weak coupling}
According to stochastic thermodynamics~\cite{ks,us},
the heat flux $\dot{q}_i$ flowing  from the $i$-th oscillator into the environment is given 
as the product of the exerted force on the oscillator and its kinematic velocity
as $\dot{q}_i\equiv l {\mathcal F}_i \circ l\dot{\phi}_i$, where $\circ$ denotes the Stratonovich product.
Then the total energy dissipation rate $P$ can be calculated as the sum of the noise average of $\dot{q}_i$ as
\begin{eqnarray}
P&&\equiv \sum_{i=1}^2 \left<\dot{q}_i\right>=l^2 \sum_{i=1}^2 \int_0^{2\pi}d\phi_1 \int_0^{2\pi}d\phi_2 {\mathcal F}_i {\mathcal J}_i^{\rm ss}(\phi_1, \phi_2)\nonumber\\
&&=\gamma l^2 \sum_{i=1}^2 \int_0^{2\pi}d\phi_1 \int_0^{2\pi}d\phi_2 (\omega_i+K\Gamma_{\rm e}(\phi_1-\phi_2))\nonumber\\ 
&& \times {\mathcal J}_i^{\rm ss}(\phi_1,\phi_2),\label{eq.power}
\end{eqnarray}
where we defined the drag coefficient $\gamma$ as $\gamma \equiv \mu^{-1}$, and used $\bigl[U{\mathcal J}_i^{\rm ss} \bigr]_0^{2\pi}=0$ because of the periodicity of $U$ and ${\mathcal J}_i^{\rm ss}$ 
and the stationarity $\frac{\partial p^{\rm ss}(\phi_1,\phi_2)}{\partial t}=0$ in Eq.~(\ref{fp.eq}). 
Under the weak coupling condition $K \ll |\omega_i|$, 
we can simplify Eq.~(\ref{eq.power}) as
$P=\gamma l^2\sum_{i=1}^2 \omega_i \Omega_i+2\gamma l^2 \bar{\omega}K\left<\Gamma_{\rm e}(\theta)\right>+O(K^2)$ 
where the small quantity of $O(K^2)$ arises from $\Gamma_{\rm e}$. For $\Gamma=\Gamma_{\rm o}$, it vanishes as well as the second term.
By using Eqs.~(\ref{eq.omega_center}) and (\ref{eq.omega_relative}), and neglecting the term of $O(K^2)$,
we obtain $P$ in a form that highlights the role of the odd and even coupling functions $\Gamma_{\rm o}$ and $\Gamma_{\rm e}$ as
\begin{eqnarray}
P=\gamma l^2(\omega_1^2+\omega_2^2)+P_{\rm o}+P_{\rm e},\label{eq.main_result}
\end{eqnarray}
where $P_{\rm o}$ and $P_{\rm e}$ are defined by
\begin{eqnarray}
&&P_{\rm o}\equiv \gamma l^2 K\Delta \omega \left<\Gamma_{\rm o}(\theta)\right>=\gamma l^2 \frac{\Delta \omega(\Delta \Omega-\Delta \omega)}{2},\label{eq.p_odd}\\
&&P_{\rm e}\equiv 4\gamma l^2 \bar{\omega}K\left<\Gamma_{\rm e}(\theta)\right>=4\gamma l^2 \bar{\omega}(\bar{\Omega}-\bar{\omega}),\label{eq.p_even}
\end{eqnarray}
respectively.
The former stems from $\Gamma_{\rm o}$ and 
depends on the frequency tuning because of coupling, whereas the latter stems from $\Gamma_{\rm e}$ and depends on the phase difference. 
In the limit of $K\to 0$ ($\Delta \Omega=\Delta \omega$), 
$P=\gamma l^2(\omega_1^2+\omega_2^2)$ 
follows from two uncoupled oscillators.
The concise and general expression of the energy dissipation rate Eq.~(\ref{eq.main_result})
for the coupled oscillators rotating on the circular trajectories
is the main result of the present paper.
This expression can be applied to any coupling function $\Gamma$ with any higher-order Fourier modes 
as long as the coupling strength is sufficiently small.
We can also derive the same formula for the noiseless case ($\epsilon=0$) (see Appendix~\ref{app_main_noiseless} for a derivation).

\subsection{Effect of odd part $P_{\rm o}$}
For $\Gamma=\Gamma_{\rm o}$, i.e., for a conservative force only, 
we obtain 
\begin{eqnarray}
P=\gamma l^2(\omega_1^2+\omega_2^2)+P_{\rm o}=2\gamma l^2 \bar{\omega}^2+\frac{\gamma l^2 \Delta \omega \Delta \Omega}{2}. 
\end{eqnarray}
For the frequency-synchronized state ($\Delta \Omega=0$), 
independently of the phase difference $\theta$, 
$P$ behaves as 
$P=2\gamma l^2 \bar{\omega}^2$ as if it originated from a single synchronized oscillator.
Because the second term $\gamma l^2 \Delta \omega \Delta \Omega/2$ is nonnegative due to $\mathcal{N}\ge 0$ and
$\Delta \omega \left[1-\exp \left(-\frac{\Delta \omega
\pi}{\tilde{D}}\right)\right]\ge 0$, 
dissipation always decreases by the frequency synchronization.
In the desynchronized state, dissipation increases because the coupled oscillators sometimes slip in phase unavoidably thus consuming extra energy.

\subsection{Effect of even part $P_{\rm e}$ near bifurcation point}
By contrast, for $\Gamma_{\rm e} \neq 0$, 
Eq.~(\ref{eq.main_result})
depends on the phase difference $\theta$ through $P_{\rm e}$ in Eq.~(\ref{eq.p_even}).
This can be clearly seen in a change of $P_{\rm e}$ near the bifurcation point 
$\Delta \omega = -2 K^* \Gamma_{\rm o,max} \ {\rm or} \ -2K^* \Gamma_{\rm o,min}$ for $\epsilon=0$,
where the transition between phase-slip and phase-locked states occurs.
For $\epsilon=0$, 
by expanding $\left<\Gamma_{\rm e}(\theta)\right>=\Gamma_{\rm e}(\theta)$ in $P_{\rm e}$ in Eq.~(\ref{eq.p_even}) around 
the bifurcation point $\theta^*$ as $\left<\Gamma_{\rm e}(\theta)\right>\simeq \Gamma_{\rm e}(\theta^*)+\Gamma_{\rm e}'(\theta^*)(\theta-\theta^*)$,
we obtain a square-root change of $P_{\rm e}$ for any $\Gamma$ as
\begin{eqnarray}
P_{\rm e}=P_{\rm e}^* \pm 4\gamma l^2 \bar{\omega}\Gamma_{\rm e}'(\theta^*)\sqrt{-\frac{2K^*\Gamma_{\rm o}(\theta^*)}{\Gamma_{\rm o}''(\theta^*)}}(K-K^*)^{1/2},\label{eq.square_root}
\end{eqnarray}
where $P_{\rm e}^*\equiv 4\gamma l^2 \bar{\omega}K^*\Gamma_{\rm e}(\theta^*)$ and the plus (minus)
sign corresponds to the left (right) equality in Eq.~(\ref{eq.bifurcation_range}). 
This expression is valid for any stable phase-locked state near the bifurcation point where the oscillators synchronize in out-of-phase.
Even in the presence of weak noise ($\epsilon \ne 0$),
we can expect a characteristic change of $P_{\rm e}$ reflecting Eq.~(\ref{eq.square_root}), as will be demonstrated 
in the next section.

\section{Example}\label{Example}
\subsection{Setup: Hydrodynamically coupled oscillators}
As an important example of the coupled oscillators illustrated in Fig.~\ref{machine} to which our formulation can be applied, 
we consider two rotating Stokes spheres on circular trajectories interacting through hydrodynamic coupling 
in a three-dimensional Stokes flow.
For our purpose, we here adopt the phase-description
approach of Refs.~\cite{UG2011prl,UG2012EPJE} where necessary conditions
for synchronization of active rotors with fixed trajectories by hydrodynamic coupling have been studied.
For hydrodynamically coupled Stokes spheres 
with radius $a$ moving on circles with radius $l$ whose centers 
are separated by a distance $d$ ($a\ll d$, $l\ll d$) (Fig.~\ref{oseen} (a)),
the phase evolution of the $i$-th sphere subject to noise~\cite{KDBBPSHC}
is given by (see Appendix~\ref{app_hydro_detail} for a detailed derivation)
\begin{eqnarray}
&&\dot{\phi}_1=\omega_1(\phi_1)+\gamma^{-1} l^{-1}G H(\phi_1,\phi_2)F_2(\phi_2)+\zeta_1,\label{eq.oseen_1}\\
&&\dot{\phi}_2=\omega_2(\phi_2)+\gamma^{-1} l^{-1}G H(\phi_1,\phi_2)F_1(\phi_1)+\zeta_2.\label{eq.oseen_2}
\end{eqnarray}
Here, $F_i(\phi_i)$ is the profile of the driving force to the $i$-th sphere, $\omega_i(\phi_i)\equiv F_i(\phi_i)/\gamma l$ is the phase-dependent natural frequency, 
\begin{eqnarray}
H(\phi_1,\phi_2)\equiv \cos (\phi_1-\phi_2)+\sin \phi_1 \sin \phi_2
\end{eqnarray}
is the geometric factor for the present case of the circular trajectories~\cite{UG2011prl,UG2012EPJE},
$\gamma$ is given as $\gamma=6\pi\eta a$ with $\eta$ being the viscosity by the Stokes' law,
and $G \equiv \frac{3a}{4d}$ is the hydrodynamic coupling parameter.
We here adopt
\begin{eqnarray} 
F_i(\phi_i)=F_i\left[1-A \sin 2\phi_i\right]
\end{eqnarray}
with $F_i$ being constants, 
for which the in-phase state is stable in the absence of noise and natural-frequency difference for 
$0<A<1$~\cite{UG2011prl,UG2012EPJE}. 
Displacement of the Stokes spheres with circular trajectories and the above force profile were experimentally realized in Ref.~\cite{KDBBPSHC} by using feedback-controlled optical tweezers.

These phase equations can be brought into
the form of Eqs.~(\ref{eq.phase_eq_1}) and (\ref{eq.phase_eq_2}), 
by first performing the variable transformation
\begin{eqnarray} 
\Phi_i\equiv \frac{2\pi}{T_i}\int_0^{\phi_i} \frac{d\phi'_i}{\omega_i(\phi'_i)}\label{eq.variable_transform}
\end{eqnarray}
with $T_i$ being a natural period $T_i\equiv \int_0^{T_i} dt =
\int_0^{2\pi}\frac{d\phi_i}{\omega_i(\phi_i)}$~\cite{UG2011prl,UG2012EPJE}.
The phase equations Eqs.~(\ref{eq.oseen_1}) and (\ref{eq.oseen_2}) are then rewritten as
\begin{eqnarray}
\dot{\Phi}_1=\omega_1+G\omega_1 \frac{\tilde{F}_2(\Phi_2)}{\tilde{F}_1(\Phi_1)}\tilde{H}(\Phi_1,\Phi_2)+\frac{\omega_1}{\tilde{\omega}_1(\Phi_1)}\zeta_1,\label{eq.oseen_1_vc}\\
\dot{\Phi}_2=\omega_2+G\omega_2 \frac{\tilde{F}_1(\Phi_1)}{\tilde{F}_2(\Phi_2)}\tilde{H}(\Phi_1,\Phi_2)+\frac{\omega_2}{\tilde{\omega}_2(\Phi_2)}\zeta_2,\label{eq.oseen_2_vc}
\end{eqnarray}
where we put $\omega_i=\frac{2\pi}{T_i}$ as the constant natural
frequency, $\tilde{F}_i(\Phi_i)\equiv F_i(\phi_i)$,
$\tilde{\omega}_i(\Phi_i)\equiv \omega_i(\phi_i)$, and
$\tilde{H}(\Phi_1,\Phi_2)\equiv H(\phi_1,\phi_2)$.
When $|\Delta \omega|/|\bar{\omega}|, G, \tilde{D}/|\bar{\omega}|\ll 1$, we can average Eqs.~(\ref{eq.oseen_1_vc}) and
(\ref{eq.oseen_2_vc}) over one cycle $2\pi$ as~\cite{K}
\begin{eqnarray}
\dot{\Phi}_1=\omega_1+G\bar{\omega}\Gamma(\Phi_1-\Phi_2)+{\zeta}_1,\label{eq.oseen_1_vc_ca}\\
\dot{\Phi}_2=\omega_2+G\bar{\omega}\Gamma(\Phi_2-\Phi_1)+{\zeta}_2,\label{eq.oseen_2_vc_ca}
\end{eqnarray}
to the lowest order,
where we have also assumed $|A|\ll 1$ for analytical tractability.
The coupling function is given by (see Appendix~A for details)
\begin{eqnarray}
\Gamma(\Phi_1-\Phi_2)=\frac{3}{2}\cos(\Phi_1-\Phi_2)-\frac{A}{2}\sin(\Phi_1-\Phi_2).\label{eq.cf_averaged}
\end{eqnarray}
Equations~(\ref{eq.oseen_1_vc_ca}) and (\ref{eq.oseen_2_vc_ca}) now have the same form as Eqs.~(\ref{eq.phase_eq_1}) and (\ref{eq.phase_eq_2}) with $K=G\bar{\omega}$.
The potential function of the odd part is
$U(\Phi_1-\Phi_2)=-\frac{A}{2} \cos(\Phi_1-\Phi_2)$. 
A purely odd coupling function 
was previously used in Ref.~\cite{BWPG} to model 
hydrodynamic synchronization of spatially-separated two eukaryotic flagella 
in an experimental setup (see also Refs.~\cite{GPT,GPT2}).

We note that the new phase $\Phi_i$ as introduced in Eq.~(\ref{eq.variable_transform}) slightly deviates from the actual position of the oscillator as $\phi_i$ denotes in Fig.~\ref{machine}.
However, if we neglect the small discrepancy of $O(AG,A^2)$ in the energy dissipation rate before and after the variable transformation, we can regard $l\Phi_i$ as the arc length measured from the origin to the position of the oscillator,
and after the cycle-averaging, our expression Eq.~(\ref{eq.main_result}) can be applied to the dynamics of $\Phi_i$ given by Eqs.~(\ref{eq.oseen_1_vc_ca}) and (\ref{eq.oseen_2_vc_ca}).
See Appendix~\ref{app_energy_dissi} for details.

\begin{figure}[t!]
\includegraphics[scale=0.875]{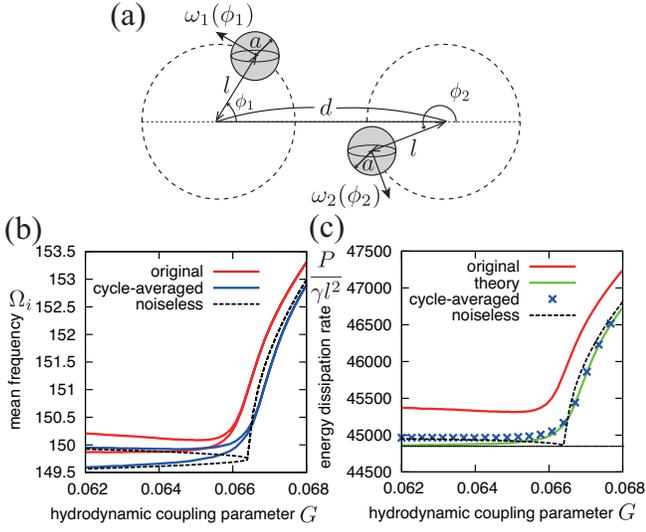}
\caption{
(a) Schematic illustration of hydrodynamically coupled Stokes spheres on circular trajectories described by Eqs.~(\ref{eq.oseen_1}) and (\ref{eq.oseen_2}). 
(b) The mean frequency $\Omega_i$ and (c) the energy dissipation rate $P$ normalized by $\gamma l^2$ as a function of the hydrodynamic coupling parameter $G$.
The numerical data obtained by Eqs.~(\ref{eq.oseen_1}) and (\ref{eq.oseen_2}) (original), Eqs.~(\ref{eq.oseen_1_vc_ca}) and (\ref{eq.oseen_2_vc_ca}) (cycle-averaged), and Eqs.~(\ref{eq.oseen_1_vc_ca}) and (\ref{eq.oseen_2_vc_ca}) with $\epsilon=0$ with other parameters being unchanged (noiseless) are compared,
where the theoretical bifurcation point $G^*=\left|\frac{\Delta
\omega}{\bar{\omega}A}\right| \simeq 0.0664$. 
The theoretical expression of $P$ in Eq.~(\ref{eq.main_result}) 
(theory) is also compared with its numerical counterpart (cycle-averaged) in (c). The dotted line in (c) denotes the contribution from the (normalized) uncoupled part 
$\omega_1^2+\omega_2^2$ in Eq.~(\ref{eq.main_result}).
}\label{oseen}
\end{figure}

\subsection{Comparison of theory with numerical calculations}
To numerically solve the equations, 
we use typical parameters for a micron-sized Stokes sphere in a viscous fluid by reference to the actual experiment~\cite{KDBBPSHC} as
$a=6.45 \mu {\rm m}$, $l=9.68 \mu {\rm m}$, $\eta=1.45 {\rm mPa}\cdot{\rm s}$, $T=300 {\rm K}$, 
$k_{\rm B}=1.38\times 10^{-23} {\rm J}{\rm K}^{-1}$, and 
$\tilde{D}=\frac{k_{\rm B}T}{\gamma l^2}=\frac{k_{\rm B}T}{6\pi \eta al^2}\simeq 2.5\times 10^{-4} {\rm s}^{-1}$.
We choose $F_1/\gamma l=150.0 {\rm s}^{-1}$ and $F_2/\gamma l=151.0 {\rm s}^{-1}$, which lead to 
$\omega_1\simeq 149.248 {\rm s}^{-1}$ and 
$\omega_2\simeq 150.243 {\rm s}^{-1}$, respectively. 
We also choose $A=0.1$ so that 
Eq.~(\ref{eq.cf_averaged}) is a good approximation.
The saddle-node bifurcation point $G^*$ for the cycle-averaged dynamics Eqs.~(\ref{eq.oseen_1_vc_ca}) and (\ref{eq.oseen_2_vc_ca}) 
in the absence of noise is determined by the combination of the three parameters as $G^*=\left|\frac{\Delta \omega}{\bar{\omega}A}\right|$ from Eq.~(\ref{eq.bifurcation_range}).  
Because in this hydrodynamic model we are assuming a regime where the cycle-averaging approximation is valid, 
we need to choose $\bar{\omega}$, $\Delta \omega$, and $G$ such that they satisfy $\left|\frac{\Delta \omega}{\bar{\omega}}\right|, G \ll 1$.
Therefore, even for given small $|A|$, we should make $G^*=\left|\frac{\Delta \omega}{\bar{\omega}A}\right|$ sufficiently small by choosing much smaller $\left|\frac{\Delta \omega}{\bar{\omega}}\right|$ than $|A|$ 
to study the energy dissipation rate around $G^*$.
The above parameters that give $G^*=\left|\frac{\Delta
\omega}{\bar{\omega}A}\right| \simeq 0.0664$ were adopted to satisfy this condition.
In the numerical calculations, the noise average is replaced with the long-time average.

Figure~\ref{oseen} (b) shows the $G$-dependence of the mean frequency
$\Omega_i$ obtained from the original dynamics given by Eqs.~(\ref{eq.oseen_1})
and (\ref{eq.oseen_2}) and the cycle-averaged one
given by Eqs.~(\ref{eq.oseen_1_vc_ca}) and (\ref{eq.oseen_2_vc_ca}) 
($\Omega_i$ obtained from the cycle-averaged dynamics Eqs.~(\ref{eq.oseen_1_vc_ca}) and (\ref{eq.oseen_2_vc_ca}) with $\epsilon=0$
is also shown as a guideline).
Although $\left<\dot{\phi}_i\right>=\left<\dot{\Phi}_i\right>$ under the variable
transformation holds for stationary states in general, a small discrepancy arises because of the cycle averaging 
that explains the bifurcation in the original dynamics at the slightly smaller $G$ than $G^*$ for the cycle-averaged dynamics.  

Figure~\ref{oseen} (c) shows the
$G$-dependence of the energy dissipation rate $P$.  The theoretical
curve is obtained by Eq.~(\ref{eq.main_result}) 
where $\Delta \Omega$ in $P_{\rm o}$ and $\bar{\Omega}$ in $P_{\rm e}$ are
derived from the data of the cycle-averaged dynamics given in Fig.~\ref{oseen} (b),
while the crosses denote the data obtained from the definition
$P=\sum_{i=1}^2 \left<\dot{q}_i\right>$ using the cycle-averaged
dynamics Eqs.~(\ref{eq.oseen_1_vc_ca}) and (\ref{eq.oseen_2_vc_ca}) (The case of $\epsilon=0$ is also shown as a guideline).
Both are in good agreement, and the discrepancy originates from the neglected term of $O(K^2)$ in Eq.~(\ref{eq.main_result}).
The original curve obtained from the definition $P=\sum_{i=1}^2
\left<\dot{q}_i\right>$ using Eqs.~(\ref{eq.oseen_1}) and
(\ref{eq.oseen_2}) is also shown for comparison.

In Fig.~\ref{oseen} (c), we can see that $P$ changes drastically around the
bifurcation point: the oscillators consume more energy as $G$ increases, i.e.,
they gain the ability to do more work on the surrounding
fluid in association with the increase of the degree of phase synchronization. 
This behavior can be explained based on Eq.~(\ref{eq.square_root}) as the effect of $\Gamma_{\rm e}$.
In the synchronized state $\Delta \Omega=0$, $P_{\rm o}$ in Eq.~(\ref{eq.p_odd}) 
becomes constant as
\begin{eqnarray}
P_{\rm o}=-\gamma l^2\frac{\Delta \omega^2}{2},\label{eq.p_odd_sync}
\end{eqnarray}
while $P_{\rm e}$ in Eq.~(\ref{eq.square_root}) with the plus sign is given as 
\begin{eqnarray}
P_{\rm e}=-6\sqrt{2}\gamma l^2\sin \theta^*\bar{\omega}^2{G^*}^{1/2}(G-G^*)^{1/2} \ \ (G\ge G^*).\label{eq.Pe_bifur}
\end{eqnarray}
See the $G$-dependence of $P_{\rm o}$ and $P_{\rm e}$ in Figs.~\ref{oseen2}(a) and~\ref{oseen2}(b), respectively, for these behaviors. 
From Eq.~(\ref{eq.Pe_bifur}), we then notice that the out-of-phase synchronization with $\theta^*=-\frac{\pi}{2}$ for $A>0$ 
gives the observed behavior, whereas the out-of-phase synchronization with $\theta^*=\frac{\pi}{2}$ for $A<0$
results in the opposite behavior; dissipation decreases as $G$ increases. 
This example elucidates an important role of the interplay between $\Gamma_{\rm o}$ and $\Gamma_{\rm e}$ on the energetics of hydrodynamic synchronization.

\red{
\begin{figure}[t!]
\includegraphics[scale=0.875]{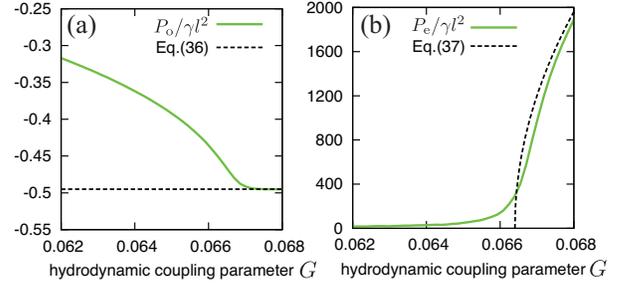}
\caption{The decomposition of the energy dissipation rate (theory) from Fig.~\ref{oseen} (c): (a) The odd part $P_{\rm o}$ as a function of the hydrodynamic coupling parameter $G$
with the (normalized) theoretical value in Eq.~(\ref{eq.p_odd_sync}). (b) The even part $P_{\rm e}$ as a function of the hydrodynamic coupling parameter $G$
with the (normalized) theoretical curve after the frequency synchronization in Eq.~(\ref{eq.Pe_bifur}) with $\theta^*=-\frac{\pi}{2}$ and $G^*=\left|\frac{\Delta \omega}{\bar{\omega}A}\right| \simeq 0.0664$. 
}\label{oseen2}
\end{figure}
}

We note that in the case of $\Delta \omega=0$ and $\epsilon=0$, we obtain 
\begin{eqnarray}
P=2\gamma l^2 \bar{\omega}^2(1+3G\cos \theta)
\end{eqnarray}
as Eq.~(\ref{eq.main_result}),
where the in-phase synchronization ($\theta=0$ for $A>0$) gives the maximum value while the anti-phase synchronization ($\theta=\pm \pi$ for $A<0$) gives the minimum value. 
Interestingly, these behaviors are opposite to those 
found in a study of hydrodynamic synchronization of two-dimensional waving sheets~\cite{EL}
where the in-phase (anti-phase) state gives a minimum (maximum) energy dissipation.
We finally stress that a measurement of $P$
via our main result 
Eq.~(\ref{eq.main_result}) is experimentally feasible without knowing the detailed $\Gamma$, 
since what is needed are only the measurable quantities of $\Omega_i$ and $\omega_i$, 
where $\omega_i$ could be measured as $\Omega_i$ for each oscillator in isolation~\cite{BWPG}.

\section{Summary and discussion}\label{Summary and discussion}
For coupled oscillators rotating on the circular trajectories described by phase equations, we have obtained a concise and general expression of the energy dissipation rate that can be applied to any type of weak coupling by using stochastic thermodynamics.
We have elucidated how
synchronization and desynchronization affect the energy dissipation rate where the decomposition of the coupling function into the even and odd parts plays the important role.
As an example, we have studied the hydrodynamic synchronization of coupled  
Stokes spheres rotating on circular trajectories in three-dimensional Stokes flow.
Although the original phase equations of this system are more complicated than
the ones we assumed in our theory, 
by using a variable transformation and cycle-averaging, we have simplified these equations into a form to which our theory can be applied. 
As predicted by our theory, these coupled Stokes spheres gain the ability to do more work on the surrounding fluid as the degree of phase synchronization increases 
under the system parameters we used.

This nonlinear dynamics feature of the energetics 
may be utilized in, e.g., propulsion of active microorganism with flagella in a viscous fluid~\cite{UG2011SM,PKT,LP}, 
where the roles of both a biochemical noise 
surpassing the thermal noise~\cite{GPT,GPT2,MKRJF} and elasticity in a complex biological environment~\cite{LP} may also become relevant issues.
In this context, 
the swimming efficiency~\cite{L} of a Stokes swimmer~\cite{FJ,PF,BG1,BG2} 
as a simple model of such propulsion 
with additional motional-degrees of freedom of a body of the microorganism 
beyond those of flagella would be worthy of further investigation.
Developing the concise description of energy dissipation for more complicated collective dynamics, 
e.g., hydrodynamic synchronization of microswimmers~\cite{PY} and 
cilia in metachronal coordination~\cite{JG,CLBJ,CLJB},  
will also be interesting.
To this end,
extensions of our theory so that it includes radial flexibility~\cite{KDBBPSHC} 
with general orbital shapes~\cite{UG2011prl,UG2012EPJE} and the formulation for many-body systems
will be required to achieve a more general formulation of the energetics of synchronization in coupled oscillators.
We expect that the present work triggers further
studies of phenomena governed by both nonequilibrium thermodynamics and nonlinear dynamics.
\begin{acknowledgements} 
The authors are grateful to Y. Nagata, M. Shigedomi, and N. Uchida for helpful discussions.
Y. I. acknowledges the financial support from a Grant-in-Aid for JSPS Fellows (Grant No. 25-9748). 
H. K. acknowledges the financial support from CREST, JST and JSPS KAKENHI Grant No. 15K16062.
The present study was supported by the JSPS Core-to-Core program ``Non-equilibrium dynamics of soft-matter and information."
\end{acknowledgements} 

\appendix

\section{Derivation of main result for noiseless case ($\epsilon=0$)}\label{app_main_noiseless}
For a derivation of the energy dissipation rate in Eq.~(\ref{eq.main_result}) in Sec.~\ref{Energy dissipation rate and synchronization} for the noiseless case ($\epsilon=0$),
we replace the Stratonovich product with the usual product and the noise average with the long-time average.
$P$ is then calculated as 
\begin{eqnarray}
P&&=\sum_{i=1}^2 \left<l \mathcal{F}_i \cdot l\dot{\phi}_i\right> \nonumber\\
&&=\sum_{i=1}^2 \lim_{T\to \infty} \frac{1}{T}\int_0^T l{\mathcal F_i}\cdot l{\dot{\phi}_i} dt\nonumber\\
&&=\gamma l^2 \sum_{i=1}^2 \lim_{T\to \infty} \frac{1}{T}\int_0^T \left(\omega_i+K\Gamma_{\rm e}(\theta)-K\frac{\partial U(\theta)}{\partial \phi_i}\right)\cdot \dot{\phi}_idt\nonumber\\
&&=\gamma l^2 \sum_{i=1}^2 \omega_i \lim_{T\to \infty}\frac{1}{T}\int_0^T \dot{\phi}_i dt \nonumber\\
&&+\gamma l^2 \sum_{i=1}^2 \lim_{T\to \infty} \frac{1}{T}\int_0^T K\Gamma_{\rm e}(\theta)\cdot (\omega_i+O(K))dt\nonumber\\
&&-\gamma l^2 K  \underbrace{\lim_{T\to \infty} \frac{1}{T} \int_{0}^{T} \sum_{i=1}^2 \frac{\partial U(\theta)}{\partial \phi_i} \frac{d\phi_i}{dt}dt}_{
=\lim_{T\to \infty} \frac{1}{T}\int_{0}^{T} \frac{dU}{dt}dt=0}\nonumber\\
&&=\gamma l^2 \sum_{i=1}^2 \omega_i \Omega_i+2\gamma l^2 \bar{\omega}K \left<\Gamma_{\rm e}(\theta)\right>+O(K^2),\label{eq.main_result_noiseless}
\end{eqnarray}
where we used $\Omega_i=\lim_{T\to \infty}\frac{1}{T}\int_0^T \dot{\phi}_i dt$, $\left<\Gamma_{\rm e}(\theta)\right>=\lim_{T\to \infty} \frac{1}{T}\int_0^T \Gamma_{\rm e}(\theta)dt$, and the fact that the potential energy $U(\theta)$ is bounded from its $2\pi$-periodicity.
By using $\bar{\Omega}=\frac{{\Omega}_1+{\Omega}_2}{2}=\bar{\omega}+K\left<\Gamma_{\rm e}(\theta)\right>$ and $\Delta \Omega=\Omega_1-\Omega_2=\Delta \omega+2K\left<\Gamma_{\rm o}(\theta)\right>$ in Eq.~(\ref{eq.main_result_noiseless}), we obtain the same expression as the main result Eq.~(\ref{eq.main_result}) 
for this noiseless case.

\section{Derivation of phase equations in hydrodynamically coupled oscillators}\label{app_hydro_detail}
We derive Eqs.~(\ref{eq.oseen_1}), (\ref{eq.oseen_2}), and (\ref{eq.oseen_1_vc_ca})--(\ref{eq.cf_averaged}) in Sec.~\ref{Example}. 
Our description is partially based on Ref.~\cite{KDBBPSHC}, where a phase-description model of hydrodynamically coupled oscillators proposed in Refs.~\cite{UG2011prl,UG2012EPJE} and its extension with radial flexibility were experimentally studied using Stokes spheres under the presence of noise.
For our purpose here, we just focus on the phase degree of freedom under the presence of noise by assuming that the radial flexibility can be neglected.
We use basically the same notations and symbols below as in the main text.

Let us consider hydrodynamically coupled Stokes spheres with radius $a$ 
moving on circles with radius $l$ whose centers ${\bm r}_{i0}\equiv (d_i,0)$ ($i=1,2$) are separated by a distance 
$d\equiv d_2-d_1>0$
in the $x$-direction on the $x$-$y$ plane (see Fig.~2 (a) in the main text).
While we assume that these spheres are in a three-dimensional Stokes flow, 
their motions are restricted on the circles on the $x$-$y$ plane.
We define ${\bm e}_x\equiv (1, 0)$ and ${\bm e}_y\equiv (0, 1)$ 
as the unit vectors in the $x$ and $y$ directions, respectively.
Then the phase equations Eqs.~(\ref{eq.oseen_1}) and (\ref{eq.oseen_2}) in the main text of the $i$-th sphere at position 
${\bm r}_i={\bm r}_{i0}+l\cos \phi_i{\bm e}_x+l\sin \phi_i{\bm e}_y=(d_i+l\cos \phi_i,l\sin \phi_i)$ can be derived from the force balance equation 
as~\cite{KDBBPSHC} 
\begin{equation}
{\bm F}_i-\sum_{j=1}^2 {\bm H}_{ij}^{-1}\cdot \dot{\bm{r}}_j+{\bm f}_i={\bm 0},\label{S1}
\end{equation}
where ${\bm F}_i$ is the profile of the driving force to the $i$-th sphere.
${\bm H}_{ij}$ is the Oseen tensor for a three-dimensional bulk fluid. 
Under the assumptions of $a\ll d$ and $l\ll d$, it is explicitly given by~\cite{KDBBPSHC}
\begin{equation}
{\bm H}_{ij}=\frac{\bm I}{\gamma}\delta_{ij}+\frac{G}{\gamma}\left({\bm I}+{\bm e}_x\otimes {\bm e}_x\right)(1-\delta_{ij}),\label{S2}
\end{equation}
where the drag coefficient $\gamma$ defined as the inverse of the mobility $\mu$ is given by the Stokes' law as $\gamma=\mu^{-1}=6\pi\eta a$ 
using the viscosity $\eta$, 
and $G=\frac{3a}{4d}$ is the hydrodynamic coupling parameter.
We define ${\bm I}\equiv {\bm e}_x\otimes {\bm e}_x+{\bm e}_y\otimes {\bm e}_y$ as the unit tensor. 
Then the inverse tensor ${\bm H}_{ij}^{-1}$ in Eq.~(\ref{S1}) is explicitly given as
\begin{widetext}
\begin{equation}
{\bm H}_{ij}^{-1}=\left(\frac{\gamma}{1-4G^2}{\bm e}_x\otimes {\bm e}_x+\frac{\gamma}{1-G^2}{\bm e}_y\otimes {\bm e}_y\right)\delta_{ij}
-\left(\frac{2\gamma G}{1-4G^2}{\bm e}_x\otimes {\bm e}_x+\frac{\gamma G}{1-G^2}{\bm e}_y\otimes {\bm e}_y\right)(1-\delta_{ij}).
\label{S3}
\end{equation} 
\end{widetext}
We denote by ${\bm f}_i(t)$ the thermal random force that satisfies $\left<{\bm f}_i(t)\right>=\bm 0$ and 
$\left<{\bm f}_i(t)\otimes{\bm f}_j(t^{\prime})\right>=2\epsilon {\bm H}_{ij}^{-1}\delta(t-t^{\prime})$ with $\epsilon$ 
being the noise intensity of the thermal environment~\cite{KDBBPSHC,DCKLC,PGQ}.
Here we are assuming that the radial degree of freedom $R_i$ of the sphere does not change in time, and that the sphere is always constrained on the circular trajectory as $R_i=l$. This assumption is justified if both the time scale of relaxation of $R_i$ to a steady value by a restoring force in the normal direction to the circle is 
much faster than the that of the phase $\phi_j$, and the radial stiffness is sufficiently large for deviation of $R_i$ from $l$ as the equilibrium point to be neglected~\cite{KDBBPSHC}. 
Under this assumption of no radial flexibility, we can assume that ${\bm F}_i$ and ${\bm f}_i$ effectively have only a tangential component as ${\bm F}_i=F_i(\phi_i){\bm t}_i$ and ${\bm f}_i=f_i{\bm t}_i$, 
where 
${\bm t}_i\equiv \frac{d{\bm r}_i}{d\phi_i}/\left|\frac{d{\bm r}_i}{d\phi_i}\right|=(-\sin \phi_i,\cos \phi_i)$
is the tangential vector to the circle.

By applying ${\bm H}_{ji}$ to Eq.~(\ref{S1}) and summing with respect to $i$, we obtain
\begin{equation}
{\bm H}_{ii}\cdot {\bm F}_i+\sum_{j\ne i} {\bm H}_{ij}\cdot {\bm
F}_j-\dot{\bm r}_i+\sum_{j=1}^2{\bm H}_{ij}\cdot{\bm f}_j={\bm 0},\label{S4}
\end{equation}
where we used $\sum_{i=1}^2{\bm H}_{ki}{\bm H}^{-1}_{ij}={\bm I}\delta_{kj}$.
By noting that the component of the velocity $\dot{{\bm r}}_i=l\dot{\phi}_i(-\sin \phi_i,\cos \phi_i)$ 
tangential to the circle is given as ${\bm t}_i\cdot \dot{{\bm r}}_i=l\dot{\phi}_i$,
we can rewrite the force balance equation Eq.~(\ref{S1}) by taking the inner product of ${\bm t}_i$ with Eq.~(\ref{S4}) multiplied by $l^{-1}$ as 
\begin{equation}
\frac{F_i(\phi_i)}{\gamma l}+\frac{G H(\phi_1,\phi_2)F_j(\phi_j)}{\gamma l}-\dot{\phi}_i+\zeta_i(t)=0\ \ \ \ (i\ne j),\label{S5}
\end{equation}
where 
$H(\phi_1,\phi_2)=\cos (\phi_1-\phi_2)+\sin \phi_1 \sin \phi_2$
is the geometric factor~\cite{UG2011prl,UG2012EPJE}, and
$\zeta_i(t)$ is defined as 
\begin{equation}
\zeta_i(t)\equiv l^{-1}\sum_{j=1}^2{\bm t}_i\cdot{\bm H}_{ij}\cdot{\bm f}_j=\frac{f_i(t)}{\gamma l}+\frac{GH(\phi_1,\phi_2)f_j(t)}{\gamma l} \ \ \ \ (i\ne j).\label{S6}
\end{equation}
To obtain the correlation function of $\zeta_i(t)$, we use  
\begin{widetext}
\begin{eqnarray}
\left<f_i(t)f_j(t^{\prime})\right>&&=2\epsilon \left({\bm t}_i\cdot {\bm H}_{ij}^{-1}\cdot{\bm t}_j\right)\delta(t-t^{\prime})\nonumber\\
&&=2\epsilon \Biggl[\left(\frac{\gamma}{1-4G^2}\sin \phi_i\sin \phi_j+\frac{\gamma}{1-G^2}\cos \phi_i\cos \phi_j\right)\delta_{ij}-\left(\frac{2\gamma G}{1-4G^2}\sin \phi_i\sin \phi_j+\frac{\gamma G}{1-G^2}\cos \phi_i\cos \phi_j\right)(1-\delta_{ij})\Biggr]\delta(t-t')\nonumber\\
&&=2\gamma \epsilon \left(\delta_{ij}-G H(\phi_1,\phi_2)(1-\delta_{ij})\right)\delta(t-t')+O(\epsilon G^2).\label{S7}
\end{eqnarray}
\end{widetext}
We then approximate $\zeta_i$ as the independent Gaussian white noise whose correlation function is given as 
$\left<\zeta_i(t)\zeta_j(t^{\prime})\right>=2\tilde{D}\delta_{ij}\delta(t-t')+O(\tilde{D}G)$,
where $\tilde D=Dl^{-2}$ is the normalized diffusion coefficient $D=\mu \epsilon$.
The force balance equation Eq.~(\ref{S1}) now becomes the following phase equations corresponding to Eqs.~(\ref{eq.oseen_1}) and (\ref{eq.oseen_2}) in the main text as
\begin{align}
\dot{\phi}_1=\omega_1(\phi_1)+\gamma^{-1} l^{-1}G H(\phi_1,\phi_2)F_2(\phi_2)+\zeta_1,\label{S8}\\
\dot{\phi}_2=\omega_2(\phi_2)+\gamma^{-1} l^{-1}G H(\phi_1,\phi_2)F_1(\phi_1)+\zeta_2,\label{S9}
\end{align}
where $F_i(\phi_i)=F_i\left[1-A \sin 2\phi_i\right]$ ($F_i\equiv F_{0}+\delta F_i$ ($|\delta F_i| \ll |F_0|$)) with $F_{0}$ and $\delta F_i$ being constants~\cite{UG2011prl,UG2012EPJE}. 

By the variable transformation Eq.~(\ref{eq.variable_transform}),
the phase equations Eqs.~(\ref{S8}) and (\ref{S9}) are then rewritten as
\begin{align}
\dot{\Phi}_1=\omega_1+G\omega_1 \frac{\tilde{F}_2(\Phi_2)}{\tilde{F}_1(\Phi_1)}\tilde{H}(\Phi_1,\Phi_2)+\frac{\omega_1}{\tilde{\omega}_1(\Phi_1)}\zeta_1,\label{S10}\\
\dot{\Phi}_2=\omega_2+G\omega_2 \frac{\tilde{F}_1(\Phi_1)}{\tilde{F}_2(\Phi_2)}\tilde{H}(\Phi_1,\Phi_2)+\frac{\omega_2}{\tilde{\omega}_2(\Phi_2)}\zeta_2,\label{S11}
\end{align}
which correspond to Eqs.~(\ref{eq.oseen_1_vc}) and (\ref{eq.oseen_2_vc}) in the main text.
When $|\Delta \omega|/|\bar{\omega}|, G, \tilde{D}/|\bar{\omega}|\ll 1$,
we can average Eqs.~(\ref{S10}) and
(\ref{S11}) over one cycle $2\pi$ as~\cite{K}
\begin{align}
\dot{\Phi}_1=\omega_1+G\bar{\omega}\Gamma(\Phi_1-\Phi_2)+\bar{\zeta}_1,\label{S12}\\
\dot{\Phi}_2=\omega_2+G\bar{\omega}\Gamma(\Phi_2-\Phi_1)+\bar{\zeta}_2,\label{S13}
\end{align}
to the lowest order.
The coupling function $\Gamma(\Phi_1-\Phi_2)$ regarding the phase difference is defined as
\begin{align}
\Gamma(\Phi_1-\Phi_2)\equiv \frac{1}{2\pi}\int_0^{2\pi} \frac{\tilde{F}_0(\Phi)\tilde{H}(\Phi_1-\Phi_2+\Phi, \Phi)}{\tilde{F}_0(\Phi_1-\Phi_2+\Phi)}d\Phi,\label{S14}
\end{align} 
where $\tilde{F}_0(\Phi_i)\equiv {F}_0(\phi_i)=F_0\left[1-A \sin 2\phi_i \right]$.
$\bar{\zeta}_i$ is the Gaussian white noise that satisfies $\left<\bar{\zeta}_i(t)\right>=0$ and $\left<\bar{\zeta}_i(t)\bar{\zeta}_j(t')\right>=2\bar{D} \delta_{ij}\delta(t-t')$, where 
$\bar{D}$ is the cycle-averaged diffusion coefficient as~\cite{K} 
\begin{equation}
\bar{D} \equiv \frac{\tilde{D}}{2\pi}\int_0^{2\pi}\frac{{\bar{\omega}}^2}{\tilde{\omega}_0^2(\Phi_i)}d\Phi_i,\label{S15}
\end{equation}
where $\tilde{\omega}_0(\Phi_i)\equiv \tilde{F}_0(\Phi_i)/\gamma l$. 

By assuming $|A| \ll 1$ for analytical tractability~\cite{UG2012EPJE}, we can approximate $\phi_i\simeq \Phi_i+\frac{A}{2}\cos 2\Phi_i=\Phi_i+O(A)$, and hence $\tilde{F}_0(\Phi_i)=F_0[1-A\sin 2\Phi_i]+O(A^2)$, 
$\tilde{H}(\Phi_1,\Phi_2)=H(\Phi_1,\Phi_2)+A\left(\sin \Phi_1\cos \Phi_2\cos 2\Phi_2+\cos \Phi_1 \sin \Phi_2 \cos 2\Phi_1\right)-\frac{A}{2}\left(\sin \Phi_1\cos \Phi_2\cos 2\Phi_1+\cos \Phi_1\sin \Phi_2\cos 2\Phi_2\right)+O(A^2)$, and $\bar{D}=
\tilde{D}(1+O(A^2))$.
With this approximation, we can reduce Eqs.~(\ref{S12}) and (\ref{S13}) to
\begin{align}
\dot{\Phi}_1=\omega_1+G\bar{\omega}\Gamma(\Phi_1-\Phi_2)+\zeta_1,\label{S16}\\
\dot{\Phi}_2=\omega_2+G\bar{\omega}\Gamma(\Phi_2-\Phi_1)+\zeta_2,\label{S17}
\end{align}
by neglecting the quantity of $O(GA^2,\tilde{D}A^2)$. Here, the coupling function Eq.~(\ref{S14}) is calculated as 
\begin{align}
\Gamma(\Phi_1-\Phi_2)=\frac{3}{2}\cos(\Phi_1-\Phi_2)-\frac{A}{2}\sin(\Phi_1-\Phi_2),
\label{S18}
\end{align} 
up to $O(A)$, which corresponds to Eq.~(\ref{eq.cf_averaged}) in the main text.
Equations~(\ref{S16}) and (\ref{S17}) correspond to Eqs.~(\ref{eq.oseen_1_vc_ca}) and (\ref{eq.oseen_2_vc_ca}) in the main text.

\section{Energy dissipation rate under variable transformation}\label{app_energy_dissi}
We show that the energy dissipation rate $P$ obtained from the original dynamics 
Eqs.~(\ref{eq.oseen_1}) and (\ref{eq.oseen_2}) with the variable $\phi_i$ 
can be rewritten by using $\Phi_i$ with the dynamics Eqs.~(\ref{eq.oseen_1_vc}) and (\ref{eq.oseen_2_vc})
via the relation 
\begin{eqnarray}
\dot{\phi}_i=\frac{\tilde{\omega}_i(\Phi_i)}{\omega_i}\dot{\Phi}_i=\frac{T_i}{2\pi}\tilde{\omega}_i(\Phi_i)\dot{\Phi}_i\label{eq.variable_transform_derivative}
\end{eqnarray}
obtained from Eq.~(\ref{eq.variable_transform}).
For this purpose,
we rewrite the original phase equations Eqs.~(\ref{eq.oseen_1}) and (\ref{eq.oseen_2}) for $\phi_i$ as 
\begin{eqnarray}
\dot{\phi}_i=\omega_i(\phi_i)+G\Gamma_i(\phi_1,\phi_2)+\zeta_i\equiv \mu {\mathcal F}_i(\phi_1,\phi_2)+\zeta_i,\label{eq.original_rewrite}
\end{eqnarray}
where we put
\begin{eqnarray}
&&\Gamma_1 (\phi_1,\phi_2)\equiv H(\phi_1, \phi_2)\frac{F_2(\phi_2)}{\gamma l},\label{eq.original_rewrite_1}\\
&&\Gamma_2 (\phi_1,\phi_2)\equiv H(\phi_1, \phi_2)\frac{F_1(\phi_1)}{\gamma l},\label{eq.original_rewrite_2}
\end{eqnarray}
respectively.
Then we can also rewrite Eqs.~(\ref{eq.oseen_1_vc}) and (\ref{eq.oseen_2_vc}) for $\Phi_i$ as
\begin{eqnarray}
\dot{\Phi}_i=\mu \frac{\omega_i\tilde{\mathcal F}_i(\Phi_1,\Phi_2)}{\tilde{\omega}_i(\Phi_i)}+\frac{\omega_i}{\tilde{\omega_i}(\Phi_i)}\zeta_i,\label{eq.variable_transformed_eq_rewrite}
\end{eqnarray}
where 
\begin{eqnarray}
\tilde{\mathcal F}_i(\Phi_1,\Phi_2)\equiv {\mathcal F}_i(\phi_1,\phi_2).
\label{eq.tilde_force_approx}
\end{eqnarray}
We also use the following approximations:
\begin{eqnarray}
T_i&&=\int_0^{2\pi}\frac{d\phi_i}{\omega_i(\phi_i)}=\int_0^{2\pi}\frac{1}{\frac{F_i}{\gamma l}\left[1-A\sin 2\phi_i\right]}d\phi_i \nonumber\\ 
&&=\frac{2\pi \gamma l}{F_i}+O(A^2),\label{eq.period_approx}\\
\tilde{\omega}_i(\Phi_i)&&=\frac{F_i}{\gamma l} \left[1-A\sin 2\Phi_i\right]+O(A^2).\label{eq.tilde_omega_approx}
\end{eqnarray}
In the following, we consider the noiseless case ($\epsilon=0$) and the case under the presence of noise ($\epsilon \ne 0$), respectively.
\subsection{Noiseless case ($\epsilon=0$)}
By using Eqs.~(\ref{eq.variable_transform_derivative}), Eq.~(\ref{eq.variable_transformed_eq_rewrite}) with $\epsilon=0$, and Eqs.~(\ref{eq.tilde_force_approx})--(\ref{eq.tilde_omega_approx}), we obtain
\begin{eqnarray}
P&&=\sum_{i=1}^2 \left<l \mathcal{F}_i(\phi_1,\phi_2) \cdot l\dot{\phi}_i\right> \nonumber\\
&&=\sum_{i=1}^2 \lim_{T\to \infty} \frac{1}{T}\int_0^T l{\mathcal F_i}(\phi_1,\phi_2)\cdot l{\dot{\phi}_i} dt\nonumber\\
&&=\sum_{i=1}^2 \lim_{T\to \infty} \frac{1}{T}\int_0^T \left(l \frac{\omega_i\tilde{\mathcal F}_i(\Phi_1,\Phi_2)}{\tilde{\omega}_i(\Phi_i)}\frac{\tilde{\omega}_i(\Phi_i)}{\omega_i}\right) \cdot \left(l \frac{\tilde{\omega}_i(\Phi_i)}{\omega_i}{\dot{\Phi}_i}\right) dt\nonumber\\
&&=l^2 \sum_{i=1}^2 \lim_{T\to \infty} \frac{1}{T}\int_0^T \frac{\omega_i\tilde{\mathcal F}_i(\Phi_1,\Phi_2)}{\tilde{\omega}_i(\Phi_i)}\cdot  \left(\frac{1}{2\pi}\right)^2\left(\frac{2\pi \gamma l}{F_i}+O(A^2)\right)^2 \nonumber\\
&&\times \left(\frac{F_i}{\gamma l} \left[1-A\sin 2\Phi_i\right]+O(A^2)\right)^2 {\dot{\Phi}_i} dt\nonumber\\
&&=l^2 \sum_{i=1}^2 \lim_{T\to \infty} \frac{1}{T}\int_0^T  \frac{\omega_i\tilde{\mathcal F}_i(\Phi_1,\Phi_2)}{\tilde{\omega}_i(\Phi_i)}  \left[1-2A\sin 2\Phi_i\right] {\dot{\Phi}_i}dt\nonumber\\
&&+O(A^2)\label{eq.energy_dissi_variable_transform1}\\
&&=l^2 \sum_{i=1}^2 \lim_{T\to \infty} \frac{1}{T}\int_0^T  \frac{\omega_i\tilde{\mathcal F}_i(\Phi_1,\Phi_2)}{\tilde{\omega}_i(\Phi_i)}\cdot {\dot{\Phi}_i} dt\nonumber\\
&&-2\gamma l^2 A \sum_{i=1}^2 \omega_i \underbrace{\lim_{T\to \infty} \frac{1}{T}\int_{\Phi_i(0)}^{\Phi_i(T)} \sin 2\Phi_id\Phi_i}_{\lim_{T\to \infty}\frac{1}{T}\left[-\frac{\cos 2\Phi_i}{2}\right]_{\Phi_i(0)}^{\Phi_i(T)}=0}+O(AG,A^2)\nonumber\\
&&=\sum_{i=1}^2 \left<l \frac{\omega_i\tilde{\mathcal F}_i(\Phi_1,\Phi_2)}{\tilde{\omega}_i(\Phi_i)}\cdot l{\dot{\Phi}_i}\right>+O(AG,A^2).\label{eq.app_dissi_transform}
\label{eq.energy_dissi_variable_transform2}
\end{eqnarray}
This form of the first term in Eq.~(\ref{eq.energy_dissi_variable_transform2}) together with the dynamics Eq.~(\ref{eq.variable_transformed_eq_rewrite}) with $\epsilon=0$
allows us to interpret that $l\Phi_i$ effectively denotes the arc length measured from the origin to the position of the oscillator in this noiseless case.
Our main result Eq.~(\ref{eq.main_result}) as is also shown for this noiseless case in Appendix~\ref{app_main_noiseless} can then be applied to this form after the suitable cycle averaging of Eqs.~(\ref{eq.oseen_1_vc}) and (\ref{eq.oseen_2_vc}) with $\epsilon=0$ 
into the form of Eqs.~(\ref{eq.oseen_1_vc_ca}) and (\ref{eq.oseen_2_vc_ca}).
\subsection{Under the presence of noise ($\epsilon \ne 0$)}
By replacing the long-time average with the noise average in Eq.~(\ref{eq.energy_dissi_variable_transform1}), we can obtain the same expression as Eq.~(\ref{eq.app_dissi_transform}) under the presence of noise as
\begin{eqnarray}
P&&=\sum_{i=1}^2 \left<l{\mathcal F_i}(\phi_1,\phi_2)\circ l\dot{\phi}_i \right>\nonumber\\
&&=l^2 \sum_{i=1}^2 \left<\frac{\omega_i\tilde{\mathcal F}_i(\Phi_1,\Phi_2)}{\tilde{\omega}_i(\Phi_i)}  \left[1-2A\sin 2\Phi_i\right] \circ {\dot{\Phi}_i}  \right>+O(A^2)\nonumber\\
&&=l^2 \sum_{i=1}^2 \left<\frac{\omega_i\tilde{\mathcal F}_i(\Phi_1,\Phi_2)}{\tilde{\omega}_i(\Phi_i)}\circ {\dot{\Phi}_i}\right>-2\gamma l^2 A\sum_{i=1}^2 \omega_i 
\left<\sin 2\Phi_i \circ \dot{\Phi}_i\right>\nonumber\\
&&+O(AG,A^2)\nonumber\\
&&=\sum_{i=1}^2 \left<l \frac{\omega_i\tilde{\mathcal F}_i(\Phi_1,\Phi_2)}{\tilde{\omega}_i(\Phi_i)}\circ l{\dot{\Phi}_i}\right>+O(AG,A^2),\label{eq.energy_dissi_variable_transform3}
\end{eqnarray}
where we used 
$\left<\sin 2\Phi_i \circ \dot{\Phi}_i\right>=0$. This can be shown as
\begin{widetext}
\begin{eqnarray}
\left<\sin 2\Phi_i \circ \dot{\Phi}_i\right>&&=\int_0^{2\pi}d\Phi_1 \int_0^{2\pi} d\Phi_2 \sin 2\Phi_i \tilde{{\mathcal{J}}}_i^{\rm ss}(\Phi_1,\Phi_2)\nonumber\\
&&=\int_0^{2\pi}d\Phi_1 \int_0^{2\pi} d\Phi_2 \frac{\partial}{\partial \Phi_i}\left(-\frac{1}{2}\cos 2\Phi_i\right) \tilde{{\mathcal{J}}}_i^{\rm ss}(\Phi_1,\Phi_2) \nonumber\\
&&=\underbrace{\int_0^{2\pi}\left[-\frac{1}{2}\cos 2\Phi_i \tilde{{\mathcal{J}}}_i^{\rm ss}(\Phi_1,\Phi_2) \right]_{\phi_i=0}^{\phi_i=2\pi}d\Phi_j}_{=0}
-\int_0^{2\pi}d\Phi_1 \int_0^{2\pi} d\Phi_2\left(-\frac{1}{2}\cos 2\Phi_i\right) \underbrace{\frac{\partial \tilde{{\mathcal{J}}}_i^{\rm ss}(\Phi_1,\Phi_2)}{\partial \Phi_i}}_{=-\frac{\partial \tilde{{\mathcal{J}}}_j^{\rm ss}(\Phi_1,\Phi_2)}{\partial \Phi_j}} \ \ \ (i\ne j) \nonumber\\
&&=\int_0^{2\pi}d\Phi_i \left(-\frac{1}{2}\cos 2\Phi_i\right)\underbrace{\left[{\tilde{\mathcal{J}}}_j^{\rm ss}(\Phi_1,\Phi_2)\right]_{\phi_j=0}^{\phi_j=2\pi}}_{=0}\nonumber\\
&&=0,
\end{eqnarray}
\end{widetext}
where we used the stationary solution of the Fokker-Planck equation for $\Phi_i$ corresponding to the dynamics Eqs.~(\ref{eq.oseen_1_vc}) and (\ref{eq.oseen_2_vc}) in the main text with $\tilde{\mathcal{J}}_i^{\rm ss}(\Phi_1,\Phi_2)$ being its stationary probability current.


\begin{references}
\bibitem{note1} In general, there may be multiple stable phase-locked solutions 
in the range given by Eq.~(\ref{eq.bifurcation_range}) when $\Gamma_{\rm o}$ has more than one pair of extrema.
As $K$ varies, such a stable solution disappears via
a saddle-node bifurcation at $\Delta \omega=-2K\Gamma_{\rm o,extrema}$, where $\Gamma_{\rm o, extrema}$ is one of the extrema
of $\Gamma_{\rm o}$. Our main result Eq.~(\ref{eq.main_result}) can also be applied to these cases.
\end{references}
\end{document}